\documentclass[sigplan,screen]{acmart}

\AtBeginDocument{%
  }

\usepackage{xspace}
\usepackage{graphicx}
\usepackage{hyperref}
\usepackage{mathtools}

\definecolor{superlightgray}{gray}{0.90}
\newcommand{\command}[1]{\begin{color}{blue}\emph{#1}\end{color}}
\newcommand{\aggregate}[1]{\begin{color}{red}\emph{#1}\end{color}}
\newcommand{\event}[1]{\begin{color}{orange}\emph{#1}\end{color}}
\newcommand{\policy}[1]{\begin{color}{violet}\emph{#1}\end{color}}
\newcommand{\projection}[1]{\begin{color}{green}\emph{#1}\end{color}}
\newcommand{\view}[1]{\begin{color}{brown}\emph{#1}\end{color}}

\newcommand{\Crem}{\emph{Crem}\xspace}

\newcommand{\preprime}[1]{\prescript{\prime}{}{#1}{}{}}

\graphicspath{ {./images/} }

\newcommand{\keyclass}   {\normalfont{\textbf{class}}}
\newcommand{\keydata}    {\normalfont{\textbf{data}}}
\newcommand{\keyforall}  {\normalfont{\textbf{forall}}}
\newcommand{\keyinstance}{\normalfont{\textbf{instance}}}

\newcommand{\keycase}    {\normalfont{\textbf{case}}}
\newcommand{\keyof}      {\normalfont{\textbf{of}}}
\newcommand{\keynewtype} {\normalfont{\textbf{newtype}}}
\newcommand{\keywhere}   {\normalfont{\textbf{where}}}

\newcommand{\variable}[1]{$\mathit{#1}$}

\newcommand{\tripleplus}{\ensuremath{\mathbin{+\mkern-5mu+\mkern-5mu+}}}
\newcommand{\triplebar}{\ensuremath{\mathbin{|\mkern0mu|\mkern0mu|}}}

\usepackage{balance}

\begin{document}

%%
%% The "title" command has an optional parameter,
%% allowing the author to define a "short title" to be used in page headers.
\title[CREM]{Cr\`eme de la Crem: Composable Representable Executable Machines}
\subtitle{Architectural Pearl}

\author{Marco Perone}
\affiliation{%
  \city{Treviso}
  \country{Italy}
}
\email{pasafama@gmail.com}
\orcid{0000-0002-1004-0431}

\author{Georgios Karachalias}
\affiliation{%
  \institution{Tweag}
  \city{Paris}
  \country{France}
}
\email{georgios.karachalias@tweag.io}
\orcid{0009-0008-3071-7842}

%%
%% By default, the full list of authors will be used in the page
%% headers. Often, this list is too long, and will overlap
%% other information printed in the page headers. This command allows
%% the author to define a more concise list
%% of authors' names for this purpose.
\renewcommand{\shortauthors}{Perone and Karachalias}

%%
%% The abstract is a short summary of the work to be presented in the
%% article.
\begin{abstract}
In this paper we describe how to build software architectures as a composition of state machines, using ideas and principles from the field of Domain-Driven Design. By definition, our approach is \emph{modular}, allowing one to compose independent subcomponents to create bigger systems, and \emph{representable}, allowing the implementation of a system to be kept in sync with its graphical representation.

In addition to the design itself we introduce the \Crem library, which provides a concrete state machine implementation that is both compositional and representable. \Crem uses Haskell's advanced type-level features to allow users to specify allowed and forbidden state transitions, and to encode complex state machine---and therefore domain-specific---properties. Moreover, since \Crem's state machines are representable, \Crem can automatically generate graphical representations of systems from their domain implementations.
\end{abstract}

%%
%% The code below is generated by the tool at http://dl.acm.org/ccs.cfm.
%% Please copy and paste the code instead of the example below.
%%


%%
%% Keywords. The author(s) should pick words that accurately describe
%% the work being presented. Separate the keywords with commas.

% \received{20 February 2007}
% \received[revised]{12 March 2009}
% \received[accepted]{5 June 2009}

%%
%% This command processes the author and affiliation and title
%% information and builds the first part of the formatted document.
\maketitle

% =============================================================================
\section{Introduction}

Tactical Domain-Driven Design focuses on identifying transactional boundaries and is often combined with an event-based architecture~\citep{Stopford:Event-Driven}, which is based on the flow of messages (e.g. commands and events) throughout an application domain and the clear separation of responsibilities among several components (e.g. aggregates, policies and projections).

Such separation often allows to fruitfully discuss with non-technical domain experts the details of the inner workings of the domain itself and to translate them directly into working code. Moreover, as the understanding of the domain deepens as time passes, the architecture itself allows for refactorings towards deeper insights.

Still, terms and concepts like aggregates, policies and projections lack a precise definition and delimitation, and this tends to create discussions and confusion.

Another issue we experienced developing systems using Domain-Driven Design techniques is the distance which appears between the theoretical model developed through distilling knowledge from the domain experts and the concrete implementation of such a model. This is basically the same issue which happens with stale documentation, where the documentation of a piece of code is not up-to-date with the current behaviour of the code itself.

With the work described in this paper we try to mitigate these issues bringing together ideas from Domain-Driven Design, state machines and functional programming, expressing DDD and event-based architectures in terms of state machines using Haskell.

The main novel contributions of this paper are:

\begin{itemize}
  \item
    Using state machines to implement policies and projections, building on top of the existing knowledge of treating aggregates as state machines~\citep{Ploch:Aggregate}.

    This allows to use state machines as the unique underlying concept needed to implement a whole application domain.

    Moreover, such an approach makes it all compositional, since state machines compose extremely well.
  \item
    Implementing state machines, and therefore architectures based on them, in such a way that information could be extracted from the implementation and used as documentation of the system.

    Specifically, we are able to generate a graphical representation out of the domain implementation which describes the relevant information about the domain itself. This is extremely helpful when discussing the behaviour of the domain with domain experts, removing the need for understanding and explaining the bare code behaviour.

    The ability of generating a graphical representation out of a state machine was already available with \emph{motor}~\citep{Wickstrom:Motor}. While \emph{motor} is focused on building single state machines, this article and the \Crem library improve the situation introducing compositionality.
  \item
    Making explicit the connection between an abstract model and its concrete implementation via a novel combination of Domain-Driven Design ideas and state machines.
  \item
    Using Haskell's powerful type system to encode complex properties of the implemented state machines, and therefore of the domains they represent.

    With \Crem we are able to express explicitly which state transitions are allowed and which, on the other hand, are forbidden.

    Using Haskell we are able to get access both to features mimicking dependent types, needed for the whole machinery to work, and also to the whole ecosystem of production-ready libraries and frameworks.

\end{itemize}

For example, with \Crem we can create systems, with an architecture based on Domain-Driven Design principles, just by composing state machines as follows
\[
\begin{array}{@{\hspace{0mm}}l@{\hspace{0mm}}}
  \mathit{wholeCartDomain} :: \mathit{StateMachine}~\mathit{CartCommand}~[\mathit{CartView}] \\
  \mathit{wholeCartDomain} = \mathit{Kleisli} \\
  \quad (\mathit{Feedback}~\mathit{cart}~\mathit{paymentGateway}) \\
  \quad \mathit{paymentStatus} \\
\end{array}
\]
and generate a graphical representation describing the implemented system (see Figure~\ref{fig:cart_payment}).

% =============================================================================
\section{Domain-Driven Design}

Domain-driven design, as a community and as a practice, aims to build software for complex domains creating a shared understanding of the domain and expressing it in a model, which describes the problem space~\citep{Verraes:DDD}.

\begin{figure}[t]
\includegraphics[width=\columnwidth]{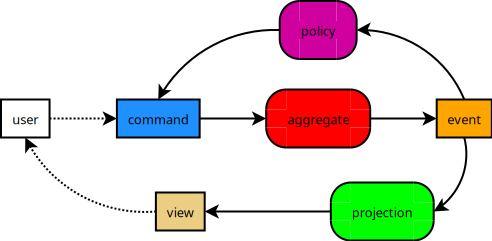}
\caption{Domain-Driven Design Model.
Arrow direction denotes the flow on information.
Solid arrows describe software interactions, while dotted ones refer to human ones.
Square boxes represent data types, while rounded ones represent processes.}
\label{fig:ddd}
\end{figure}

One popular shape of such models, which emerges naturally from Event Storming workshops~\citep{Brandolini:Event-Storming}, is rooted in ideas such as Aggregates~\citep{Evans:DDD}, Event Sourcing~\citep{Young:ES}\footnote{Notice that the architecture that we describe in the paper is very well suited for Event Sourcing, but does not rely on it and is actually independent of how persistence is dealt with} and CQRS~\citep{Young:CQRS} and is captured in Figure~\ref{fig:ddd}.

The flow of such an architecture goes as follows:
\begin{itemize}
  \item
    A user expresses their desired interaction with the software with a \command{Command}, comprising all the necessary information to execute that command.
  \item
    The \command{Command} is received by an \aggregate{Aggregate}, which has the role of making sure that every invariant of the system is preserved. Then, it emits its decisions as \event{Events}.
  \item
    \event{Events} describe the relevant state transitions of the system, containing all the data necessary to propagate information through the system and potentially to reconstruct the current state of the system.
  \item
    \policy{Policies} describe the reactive logic of the application. Whenever an \event{Event} is emitted by an \aggregate{Aggregate}, the system might decide to react to it emitting a new \command{Command}.
  \item
    \projection{Projections} aggregate the information contained in the \event{Events} and condensate them into specialised \view{Views}.
  \item
    \view{Views} describe the information which are then shown to and used by the user to decide which \command{Command} they should require next.
\end{itemize}

We will henceforth use the colors used for defining the terms whenever there is an instance of this term. For example, as
  \aggregate{Aggregate} is colored red\footnote{Traditionally in \variable{Event Storming} aggregates are yellow, but it doesn't render well on paper, so we switched to red.}, the \aggregate{Cart} aggregate will also be red.

As a recurring example, we will consider the checkout process of a standard ecommerce systems. In such a context, a user emits a \command{PayCart} command instructing the system to process the payment of their cart. The \aggregate{Cart} aggregate will check all the system invariants to ensure that the cart is in a state where it could actually be paid (e.g. it is not empty); if every invariant is satisfied, the aggregate will emit a \event{CartPaymentInitiated} event. A \policy{PaymentGateway} policy will contact an external system to actually process the payment; if the payment is processed correctly, the policy will emit a \command{MarkCartAsPaid} command. The command will then be processed by the \aggregate{Cart} aggregate, which will emit a \event{CartPaymentCompleted} event. The \projection{PaymentStatus} projection will react to the \event{CartPaymentInitiated} and \event{CartPaymentCompleted} events to provide the user with a \view{CartState} view describing the current state of the payment.

Aggregates and projections are by their nature pure stateful processes, meaning that the only effect which can take place is state management. In particular their logic does not depend on the interaction with the external world. All the relevant information for making sensible decisions is contained either in their state or provided by the incoming messages.

On the other hand, policies are by nature impure, since they often deal with interacting with external systems.

The cycle which get created between aggregates and policies helps to split the write side of the domain logic between its pure and its effectful parts, increasing the testability of the system.

% =============================================================================
\section{State machines}

In the architecture described above, while commands, events and views are just simple data, on the other hand aggregates, policies and projections are stateful processes.
As such, they could be implemented as state machines. By \emph{state machine} we always mean a \emph{Mealy machine}~\citep{Mealy}, which is composed of a stateful function and the current value for the state. Practically, it consists of the current value of the state and an action to emit an output determined from an input and the current value of the state, while being able to update the state.

In Haskell terms, one potential implementation could be the following.
\[
\begin{array}{c@{\hspace{1mm}}l@{\hspace{1mm}}l@{\hspace{1mm}}l}
  \multicolumn{4}{l}{\keydata~\mathit{Mealy}~s~a~b = \mathit{Mealy}} \\
  \quad\{ & \mathit{initialState} & :: & s\\
  \quad,  & \mathit{action}       & :: & s \to a \to (b, s)\\
  \quad\} \\
\end{array}
\]
State machines come with many practical benefits:
\begin{itemize}
  \item
    They are extremely compositional and allow combining simple state machines into more complex ones in several ways.

    For examples, two machines could be composed sequentially, feeding the output of the first machines as inputs of the second.

    Or they could be executed in parallel, providing the inputs to both machines and collecting their outputs.

    Or they could be executed in alternative, providing either the input to the first machine or the input the second, executing the corresponding machine, and obtaining as output either the output of the first machine or the output of the second.
  \item
    They allow a graphical representation as state diagrams, which could help to understand how a machine actually works also for non-technical people.
  \item
    They can be implemented as a function depending just on the \emph{input} and the \emph{state}. The simplicity of this mental model helps the implementor considering all the possible cases which need to be considered and uncovering potential edge cases.
\end{itemize}

For example, the \aggregate{Cart} aggregate of the previous section could be implemented along these lines; see Figure~\ref{fig:cart_simplified}.

\begin{figure}[t]
\[
\begin{array}{@{\hspace{0mm}}r@{\hspace{1mm}}c@{\hspace{1mm}}l}
  \multicolumn{3}{@{\hspace{0mm}}l}{\keydata~\mathit{CartCommand}} \\
  \quad & = & \mathit{PayCart} \\
  \quad & | & \mathit{MarkCartAsPaid} \\[2mm]

  \multicolumn{3}{@{\hspace{0mm}}l}{\keydata~\mathit{CartEvent}} \\
  \quad & = & \mathit{CartPaymentInitiated} \\
  \quad & | & \mathit{CartPaymentCompleted} \\[2mm]

  \multicolumn{3}{@{\hspace{0mm}}l}{\keydata~\mathit{CartState}} \\
  \quad & = & \mathit{WaitingForPayment} \\
  \quad & | & \mathit{InitiatingPayment} \\
  \quad & | & \mathit{PaymentComplete} \\[2mm]

  \multicolumn{3}{@{\hspace{0mm}}l}{\mathit{cart} :: \mathit{Mealy}~\mathit{CartState}~\mathit{CartCommand}~[\mathit{CartEvent}]} \\
  \multicolumn{3}{@{\hspace{0mm}}l}{\mathit{cart} = \mathit{Mealy}} \\
    \hspace{2mm}
      & \{ & \mathit{initialState} = \mathit{WaitingForPayment} \\
      & ,  & \mathit{action} = \backslash\keycase \\
      &    & \quad\mathit{WaitingForPayment}~\mathit{PayCart} \to \\
      &    & \qquad([\mathit{CartPaymentInitiated}], \mathit{InitiatingPayment}) \\
      &    & \quad\mathit{WaitingForPayment}~\mathit{MarkCartAsPaid} \to \\
      &    & \qquad([], \mathit{WaitingForPayment}) \\
      &    & \quad\mathit{InitiatingPayment}~\mathit{PayCart} \to \\
      &    & \qquad([], \mathit{InitiatingPayment}) \\
      &    & \quad\mathit{InitiatingPayment}~\mathit{MarkCartAsPaid} \to \\
      &    & \qquad([\mathit{CartPaymentCompleted}], \mathit{PaymentComplete}) \\
      &    & \quad\mathit{PaymentComplete}~\mathit{PayCart} \to \\
      &    & \qquad([], \mathit{PaymentComplete}) \\
      &    & \quad\mathit{PaymentComplete}~\mathit{MarkCartAsPaid} \to \\
      &    & \qquad([], \mathit{PaymentComplete}) \\
      & \} & \\
\end{array}
\]
\caption{Cart Implementation, Simplified}
\label{fig:cart_simplified}
\end{figure}

This is a very simplified version, which does not perform any sophisticated logic on duplicate or unexpected messages, but still it gives the idea of how the domain logic could be implemented.

In real systems, one will have to deal with multiple aggregates, handling different kinds of commands. Since there will be only one aggregate to deal with a given command, we can compose several aggregates in parallel to obtain a bigger component able to deal with multiple kinds of commands. Then, when we receive a command, we will need to route it to the correct component to be dealt with.

Similarly, we can compose multiple policies or projects to put ourselves back in the case when we have a single policy or projection.

% =============================================================================
\section{Crem}

The architecture we described in the Domain-Driven Design section could be implemented as a composition of state machines. Aggregates, policies and projections get implemented as state machines, which then get composed into a unique state machine which describes the whole domain of the application.

To benefit from the above described perks of both the architecture and state machines we need a way to implement a domain with state machines which is compositional and allows synchronising a graphical representation of the state machines with the actual implemented code.

\Crem is a Haskell library for defining and executing state machines in such a way that the theoretical benefits of state machines, as compositionality and representability, are practically preserved.

It allows to:
\begin{itemize}
  \item
    Compose state machines in multiple ways.
  \item
    Generate a graphical representation out of the implementation of a state machine.
  \item
    Impose invariants on the allowed transitions.
\end{itemize}

\Crem is based on two main ideas: tracking the set of the allowed transitions at the type level and using a free-like structure to restore compositionality.

Let's see how these two ideas play really well together and why both have a relevant role in the implementation.

A simple definition of a state machine that is parametric over the type of its
state $s$, as well as input and output types $a$ and $b$, respectively, could
look like the following:
\[
\begin{array}{c@{\hspace{1mm}}l@{\hspace{1mm}}l@{\hspace{1mm}}l}
  \multicolumn{4}{l}{\keydata~\mathit{Mealy}~s~a~b = \mathit{Mealy}} \\
  \quad\{ & \mathit{initialState} & :: & s\\
  \quad,  & \mathit{action}       & :: & s \to a \to (b, s)\\
  \quad\} \\
\end{array}
\]
By making the state type implicit, one ends up with $\mathit{Mealy}$ as is
defined in the \emph{machines} library~\citep{Kmett:machines}:\footnote{This transformation can be perceived as two separate steps:
\begin{enumerate}
\item
  First turn the $s$ into an existentially quantified type variable:
  \[
  \begin{array}{c@{\hspace{1mm}}l@{\hspace{1mm}}l@{\hspace{1mm}}l}
    \multicolumn{4}{l}{\keydata~\mathit{Mealy'}~a~b = \keyforall~s.~\mathit{Mealy'}} \\
    \quad\{ & \mathit{initialState} & :: & s\\
    \quad,  & \mathit{action}       & :: & s \to a \to (b, s)\\
    \quad\} \\
  \end{array}
  \]
\item
  Then use recursion to eliminate the state variable altogether (see
  $\mathit{unfoldMealy}$ from the \emph{machines} library~\citep{Kmett:machines}):
  \[
  \begin{array}{l}
    \mathit{unfoldMealy} :: \mathit{Mealy'}~a~b \to \mathit{Mealy}~a~b \\
    \mathit{unfoldMealy}~(\mathit{Mealy'}~\mathit{initial}~\mathit{action}) = \mathit{go}~\mathit{initial}~\keywhere\\
      \quad \mathit{go}~s = \mathit{Mealy} ~\$~ \backslash a \to \keycase~\mathit{action}~s~a~\keyof \\
      \qquad (b, t) \to (b, \mathit{go}~t)
  \end{array}
  \]
\end{enumerate}
}
\[
\begin{array}{c@{\hspace{1mm}}l@{\hspace{1mm}}l@{\hspace{1mm}}l}
  \multicolumn{4}{l}{\keynewtype~\mathit{Mealy}~a~b = \mathit{Mealy}} \\
  \quad\{ & \mathit{runMealy} & :: & a \to (b,~\mathit{Mealy}~a~b) \\
  \quad\} \\
\end{array}
\]
This allows improving the compositionality of the data type, since now there is no need to keep track of the state type variable, and it is possible to implement instances for common type classes like \variable{Category}, \variable{Profunctor} and \variable{Arrow}.

On the other hand, these simple implementations do not allow to extract any information about how the state machine works, without actually executing the state machine itself. Being just functions, the only thing we can do with them is running them.

In particular, we have no way to extract information about which transitions are allowed or not by the state machine, and therefore we are not able to generate a graphical representation which describes how the machine works.

To be able to enforce which state transitions are actually allowed and to generate a graphical representation of the state space, we need to track the list of allowed transitions, which we call \variable{Topology}.
\[
\begin{array}{c@{\hspace{1mm}}l@{\hspace{1mm}}l@{\hspace{1mm}}l}
  \multicolumn{4}{l}{\keynewtype~\mathit{Topology}~\mathit{vertex} = \mathit{Topology}} \\
  \quad\{ & \mathit{edges} & :: & [(\mathit{vertex}, [\mathit{vertex}])] \\
  \quad\} \\
\end{array}
\]
The \variable{Topology} is a list of edges, grouped by their initial vertex.

Coming back to our example, the \variable{Topology} of the \aggregate{Cart} aggregate could look as follows:
\[
\begin{array}{@{\hspace{0mm}}r@{\hspace{1mm}}c@{\hspace{1mm}}l}
  \multicolumn{3}{@{\hspace{0mm}}l}{\keydata~\mathit{CartVertex}} \\
    \hspace{2mm}
      & =    & \mathit{WaitingForPaymentVertex} \\
      & \mid & \mathit{InitiatingPaymentVertex} \\
      & \mid & \mathit{PaymentCompleteVertex} \\
  \\
  \multicolumn{3}{@{\hspace{0mm}}l}{\mathit{cartTopology} :: \mathit{Topology}~\mathit{CartVertex}} \\
  \multicolumn{3}{@{\hspace{0mm}}l}{\mathit{cartTopology} = } \\
    \hspace{2mm}
      & [ & (\mathit{WaitingForPaymentVertex}, [\mathit{InitiatingPaymentVertex}]) \\
      & , & (\mathit{InitiatingPaymentVertex}, [\mathit{PaymentCompleteVertex}] \\
      & , & [\mathit{PaymentCompleteVertex}, []]) \\
      & ] & \\
\end{array}
\]
By storing the \variable{Topology} at the type level we are able to use it for enforcing at compile time that forbidden transitions are never executed. Having an explicit \variable{Topology} allows to retrieve the information later on to create a graphical representation of the state machine out of it. In this way, we are sure that the generated graphical representation is always in sync with the implemented logic of the state machine.

One possible way to implement such a mechanism is depicted in Figure~\ref{fig:state_machines_revisited}.

\begin{figure}[t]
\[
\begin{array}{l}
  \keydata~\mathit{InitialState}~(\mathit{state} :: \mathit{vertex} \to \mathit{Type})~\keywhere \\
  \quad\mathit{InitialState} :: \mathit{state}~\mathit{vertex} \to \mathit{InitialState}~\mathit{state} \\[2mm]

  \keydata~\mathit{ActionResult} \\
  \qquad(\mathit{topology} :: \mathit{Topology}~\mathit{vertex}) \\
  \qquad(\mathit{state} :: \mathit{vertex} \to \mathit{Type}) \\
  \qquad(\mathit{initialVertex} :: \mathit{vertex}) \\
  \qquad\mathit{output} \\
  \quad\keywhere \\
  \quad\mathit{ActionResult} \\
  \qquad :: \mathit{AllowedTransition}~\mathit{topology}~\mathit{initialVertex}~\mathit{finalVertex} \\
  \qquad \Rightarrow (\mathit{output}, \mathit{state}~\mathit{finalVertex}) \\
  \qquad \to         \mathit{ActionResult}~\mathit{topology}~\mathit{state}~\mathit{initialVertex}~\mathit{output} \\[2mm]

  \keydata~\mathit{BaseMachine} \\
  \qquad(\mathit{topology} :: \mathit{Topology}~\mathit{vertex}) \\
  \qquad\mathit{input} \\
  \qquad\mathit{output} = \keyforall~\mathit{state}. \\
  \quad\mathit{BaseMachine} \\
  \quad\{~\mathit{initialState} :: \mathit{InitialState}~\mathit{state} \\
  \quad,~\mathit{action} \\
  \qquad ::  \keyforall~\mathit{initialVertex}. \\
  \qquad     \mathit{state}~\mathit{initialVertex} \\
  \qquad \to \mathit{input} \\
  \qquad \to \mathit{ActionResult}~\mathit{topology}~\mathit{state}~\mathit{initialVertex}~\mathit{output} \\
  \quad\}
\end{array}
\]
\caption{State Machines, Revisited}
\label{fig:state_machines_revisited}
\end{figure}

\variable{AllowedTransition}~\variable{topology}~\variable{initialVertex}~\variable{finalVertex}, which is implemented as in Figure~\ref{fig:allowed-transitions}, is a type class which checks that a transition from \variable{initialVertex} to \variable{finalVertex} is allowed by the \variable{topology}. It searches through all the edges of the \variable{Topology} whether there is one starting from \variable{initialVertex} and ending at \variable{finalVertex}, and it constructs a proof of it using the \variable{AllowTransition} data type.

\begin{figure}[t]
\[
\begin{array}{l}
  \keydata~\mathit{AllowTransition} \\
  \qquad (\mathit{topology}~::~\mathit{Topology}~\mathit{vertex}) \\
  \qquad (\mathit{initial}~::~\mathit{vertex}) \\
  \qquad (\mathit{final}~::~\mathit{vertex}) \\
  \quad \keywhere \\
  \quad \mathit{AllowIdentityEdge} \\
  \qquad ::~\mathit{AllowTransition}~\mathit{topology}~\mathit{a}~\mathit{a} \\
  \quad \mathit{AllowFirstEdge} \\
  \qquad ::~\mathit{AllowTransition}~(\preprime{\mathit{Topology}}~(\preprime{(}\mathit{a},~\mathit{b}~\preprime{:}~\mathit{l1})~\preprime{:}~\mathit{l2}))~\mathit{a}~\mathit{b} \\
  \quad \mathit{AllowAddingEdge} \\
  \qquad ::~\mathit{AllowTransition}~(\preprime{\mathit{Topology}}~(\preprime{(}\mathit{a},~\mathit{l1})~\preprime{:}~\mathit{l2}))~\mathit{a}~\mathit{b} \\
  \qquad \to~\mathit{AllowTransition}~(\preprime{\mathit{Topology}}~(\preprime{(}\mathit{a},~\mathit{x}~\preprime{:}~\mathit{l1})~\preprime{:}~\mathit{l2}))~\mathit{a}~\mathit{b} \\
  \quad \mathit{AllowAddingVertex} \\
  \qquad ::~\mathit{AllowTransition}~(\preprime{\mathit{Topology}}~\mathit{topology})~\mathit{a}~\mathit{b} \\
  \qquad \to~\mathit{AllowTransition}~(\preprime{Topology}~(\mathit{x}~\preprime{:}~\mathit{topology}))~\mathit{a}~\mathit{b}
\end{array}
\]
\[
\begin{array}{l}
  \keyclass~\mathit{AllowedTransition} \\
  \qquad (\mathit{topology}~::~\mathit{Topology}~\mathit{vertex}) \\
  \qquad (\mathit{initial}~::~\mathit{vertex}) \\
  \qquad (\mathit{final}~::~\mathit{vertex}) \\
  \quad \keywhere \\
  \quad \mathit{allowsTransition}~::~\mathit{AllowTransition}~\mathit{topology}~\mathit{initial}~\mathit{final} \\
\\
  \keyinstance \{-\#~\mathit{INCOHERENT}~\#-\} \\
  \qquad \mathit{AllowedTransition}~\mathit{topology}~\mathit{a}~\mathit{a} \\
  \quad \keywhere \\
  \quad \mathit{allowsTransition}~=~\mathit{AllowIdentityEdge} \\
\\
  \keyinstance \{-\#~\mathit{INCOHERENT}~\#-\} \\
  \qquad \mathit{AllowedTransition}~(\preprime{\mathit{Topology}}~(\preprime{(}\mathit{a},~\mathit{b}~\preprime{:}~\mathit{l1})~\preprime{:}~\mathit{l2}))~\mathit{a}~\mathit{b} \\
  \quad \keywhere \\
  \quad \mathit{allowsTransition}~=~\mathit{AllowFirstEdge} \\
\\
  \keyinstance \{-\#~\mathit{INCOHERENT}~\#-\} \\
  \qquad \mathit{AllowedTransition}~(\preprime{\mathit{Topology}}~(\preprime{(}\mathit{a},~\mathit{l1})~\preprime{:}~\mathit{l2}))~\mathit{a}~\mathit{b})~\Rightarrow \\
  \qquad \mathit{AllowedTransition}~(\preprime{\mathit{Topology}}~(\preprime{(}\mathit{a},~\mathit{x}~\preprime{:}~\mathit{l1})~\preprime{:}~\mathit{l2}))~\mathit{a}~\mathit{b} \\
  \quad \keywhere \\
  \quad \mathit{allowsTransition}~=~\mathit{AllowAddingEdge}~\mathit{allowsTransition} \\
\\
\keyinstance \{-\#~\mathit{INCOHERENT}~\#-\} \\
  \qquad \mathit{AllowedTransition}~(\preprime{\mathit{Topology}}~\mathit{topology})~\mathit{a}~\mathit{b}~\Rightarrow \\
  \qquad \mathit{AllowedTransition}~(\preprime{Topology}~(\mathit{x}~\preprime{:}~\mathit{topology}))~\mathit{a}~\mathit{b} \\
  \quad \keywhere \\
  \quad \mathit{allowsTransition}~=~\mathit{AllowAddingVertex}~\mathit{allowsTransition}\\
\end{array}
\]
\caption{\variable{AllowedTransition} implementation}
\label{fig:allowed-transitions}
\end{figure}

The \variable{InitialState} and \variable{ActionResult} data types are respectively ways to store a \variable{state} and a pair of a \variable{state} and an \variable{output}, given the fact that the state does not have kind \variable{Type} but kind $\mathit{vertex}\to\mathit{Type}$.

With these new data types, the \aggregate{Cart} aggregate could be adjusted accordingly. The updated implementation is shown in Figure~\ref{fig:cart_revisited}.

\begin{figure}[t]
\[
\begin{array}{c@{\hspace{1mm}}l@{\hspace{1mm}}l@{\hspace{1mm}}l}
  \multicolumn{4}{@{\hspace{0mm}}l}{\keydata~\mathit{CartState}~(\mathit{cartVertex}~::~\mathit{CartVertex})~\keywhere} \\
  \quad & \mathit{WaitingForPayment} & :: & \mathit{CartState}~\mathit{WaitingForPaymentVertex} \\
  \quad & \mathit{InitiatingPayment} & :: & \mathit{CartState}~\mathit{InitiatingPaymentVertex} \\
  \quad & \mathit{PaymentComplete} & :: & \mathit{CartState}~\mathit{PaymentCompleteVertex} \\
\end{array}
\]
\[
\begin{array}{@{\hspace{0mm}}r@{\hspace{1mm}}c@{\hspace{1mm}}l@{\hspace{1mm}}l}
  \multicolumn{3}{@{\hspace{0mm}}l}{\mathit{cart} :: \mathit{BaseMachine}~\mathit{CartTopology}~\mathit{CartCommand}~[\mathit{CartEvent}]} \\
  \multicolumn{3}{@{\hspace{0mm}}l}{\mathit{cart} = \mathit{BaseMachine}} \\
    \hspace{2mm}
      & \{ & \mathit{initialState} = \mathit{InitialState}~\mathit{WaitingForPayment} \\
      & ,  & \mathit{action} = \backslash\keycase \\
      &    & \hspace{1mm}\mathit{WaitingForPayment}~\mathit{PayCart} \to \\
      &    & \hspace{3mm}\mathit{ActionResult}~([\mathit{CartPaymentInitiated}], \mathit{InitiatingPayment}) \\
      &    & \hspace{1mm}\mathit{WaitingForPayment}~\mathit{MarkCartAsPaid} \to \\
      &    & \hspace{3mm}\mathit{ActionResult}~([], \mathit{WaitingForPayment}) \\
      &    & \hspace{1mm}\mathit{InitiatingPayment}~\mathit{PayCart} \to \\
      &    & \hspace{3mm}\mathit{ActionResult}~([], \mathit{InitiatingPayment}) \\
      &    & \hspace{1mm}\mathit{InitiatingPayment}~\mathit{MarkCartAsPaid} \to \\
      &    & \hspace{3mm}\mathit{ActionResult}~([\mathit{CartPaymentCompleted}], \mathit{PaymentComplete}) \\
      &    & \hspace{1mm}\mathit{PaymentComplete}~\mathit{PayCart} \to \\
      &    & \hspace{3mm}\mathit{ActionResult}~([], \mathit{PaymentComplete}) \\
      &    & \hspace{1mm}\mathit{PaymentComplete}~\mathit{MarkCartAsPaid} \to \\
      &    & \hspace{3mm}\mathit{ActionResult}~([], \mathit{PaymentComplete}) \\
      & \} & \\
\end{array}
\]
\caption{Cart Implementation, Revisited}
\label{fig:cart_revisited}
\end{figure}

Even though the implementation looks completely analogous to the previous version using the \variable{Mealy} data type, we are now ensuring at compile time that the machine will never use a transition which is not allowed by the \variable{Topology}.

The \variable{BaseMachine} data type allows us to track at the type level the information regarding the \variable{Topology} of the machine.

On the other hand, that \variable{topology} type parameter makes composition harder, since composing two machines would require computing at the type level the topology of the composed machine. Moreover, just the presence of that third type parameter does not allow us to us common type classes as \variable{Category}, \variable{Profunctor} or \variable{Arrow}.

To restore compositionality, we define a free-like structure which adds composition operations on top of the \variable{BaseMachine} data type, which is presented in Figure~\ref{fig:state_machines_final}.

\begin{figure}[t]
\[
\begin{array}{l}
  \keydata~\mathit{StateMachine}~\mathit{input}~\mathit{output}~\keywhere \\
  \quad\mathit{Basic} \\
  \qquad :: \keyforall~m~\mathit{vertex}~\mathit{topology}~\mathit{input}~\mathit{output}. \\
  \qquad\quad (~\mathit{Demote}~\mathit{vertex} \sim \mathit{vertex} \\
  \qquad\quad ,~\mathit{SingKind}~\mathit{vertex} \\
  \qquad\quad ,~\mathit{SingI}~\mathit{topology} \\
  \qquad\quad ) \\
  \qquad \Rightarrow \mathit{BaseMachine}~\mathit{topology}~\mathit{input}~\mathit{output} \\
  \qquad \to \mathit{StateMachine}~\mathit{input}~\mathit{output} \\
  \quad\mathit{Sequential} \\
  \qquad ::  \mathit{StateMachine}~a~b \\
  \qquad \to \mathit{StateMachine}~b~c \\
  \qquad \to \mathit{StateMachine}~a~c \\
  \quad\mathit{Parallel} \\
  \qquad ::  \mathit{StateMachine}~a~b \\
  \qquad \to \mathit{StateMachine}~c~d \\
  \qquad \to \mathit{StateMachine}~(a, c)~(b, d) \\
  \quad\mathit{Alternative} \\
  \qquad ::  \mathit{StateMachine}~a~b \\
  \qquad \to \mathit{StateMachine}~c~d \\
  \qquad \to \mathit{StateMachine}~(\mathit{Either}~a~c)~(\mathit{Either}~b~d) \\
  \quad\mathit{Feedback} \\
  \qquad ::  \mathit{StateMachine}~\mathit{a}~[\mathit{b}] \\
  \qquad \to \mathit{StateMachine}~\mathit{b}~[\mathit{a}] \\
  \qquad \to \mathit{StateMachine}~\mathit{a}~[\mathit{b}] \\
  \quad\mathit{Kleisli} \\
  \qquad ::  \mathit{StateMachine}~\mathit{a}~[\mathit{b}] \\
  \qquad \to \mathit{StateMachine}~\mathit{b}~[\mathit{c}] \\
  \qquad \to \mathit{StateMachine}~\mathit{a}~[\mathit{c}] \\
\end{array}
\]
\caption{State Machine Representation}
\label{fig:state_machines_final}
\end{figure}

This way, it becomes trivial to implement type classes like \variable{Category}, \variable{Strong} and \variable{Choice}, in terms of the constructors.

The \variable{Sequential} constructor allows restoring sequential categorical composition. The \variable{Parallel} allows implementing \variable{Arrow} and \variable{Strong}, while the \variable{Alternative} constructor allows implementing \variable{ArrowChoice} and \variable{Choice}.

The \variable{Feedback} constructor is used to loop two state machines, respectively feeding the output of one as input of the other.

The \variable{Kleisli} constructor allows composing sequentially two state machines which produce multiple outputs, while they process single inputs.

The \variable{StateMachine} data type helps us construct an abstract syntax tree where the leaves are \variable{BaseMachine}s and the other nodes describe how we are composing the subtrees.

Considering our running example, if we suppose now to have implemented the \aggregate{Cart} aggregate, the \policy{PaymentGateway} policy and the \projection{PaymentStatus} projection, like so
\[
\begin{array}{@{\hspace{0mm}}l@{\hspace{1mm}}c@{\hspace{1mm}}l@{\hspace{0mm}}}
  \mathit{cart}           & :: & \mathit{StateMachine}~\mathit{CartCommand}~[\mathit{CartEvent}] \\
  \mathit{paymentGateway} & :: & \mathit{StateMachine}~\mathit{CartEvent}~[\mathit{CartCommand}] \\
  \mathit{paymentStatus}  & :: & \mathit{StateMachine}~\mathit{CartEvent}~[\mathit{CartView}] \\
\end{array}
\]
we can compose them together to create an implementation for the whole domain:
\[
\begin{array}{l}
  \mathit{wholeCartDomain} :: \mathit{StateMachine}~\mathit{CartCommand}~[\mathit{CartView}] \\
  \mathit{wholeCartDomain} = \mathit{Kleisli} \\
  \quad (\mathit{Feedback}~\mathit{cart}~\mathit{paymentGateway}) \\
  \quad \mathit{paymentStatus} \\
\end{array}
\]
A whole implementation could be found in the \emph{examples} folder inside the \Crem repository~\citep{Perone:Crem}.

\begin{figure}[t]
\includegraphics[width=\columnwidth]{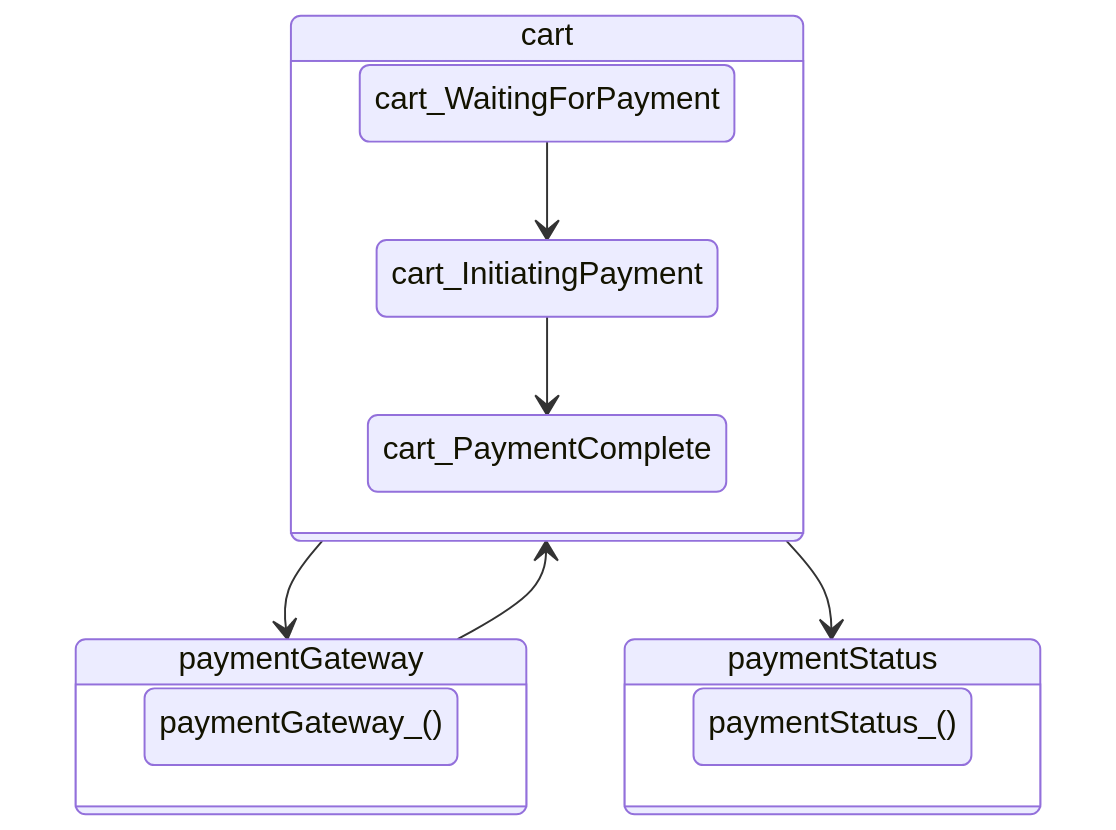}
\caption{Architecture diagram of the cart payment system}
\label{fig:cart_payment}
\end{figure}

From this definition, \Crem is able to generate a diagram that shows how the application is structured, presented in Figure~\ref{fig:cart_payment}.

From the picture, it's easy to understand how the \aggregate{cart} aggregate is connected in a loop with the \policy{paymentGateway} policy and how the outputs of the loop are then fed as inputs into the \projection{cartState} projection.

Inside every box we can also see the topology of the state space of the state machine controlling that specific component.

When we interpret the \variable{StateMachine} abstract syntax tree, we need to define how to process the leaves, i.e. the \variable{BaseMachine}s, and how to interpret the composition of substrees.

For example, when we want to run a \variable{StateMachine}, we need to describe how to run a \variable{BaseMachine}, and this is provided by the \variable{action} included in the \variable{BaseMachine} definition, and how to run a composition of sub-\variable{StateMachine}s. This depends on the specific constructor; for example, for the \variable{Sequential} constructor, we first recursively run the first \variable{StateMachine}, and we use the output as input to run the second \variable{StateMachine}.

Similarly, when we want to create a graphical representation of a \variable{StateMachine}, we need to be able first to create a graphical representation of its leaves, which are \variable{BaseMachine}s. We are able to do that because we have the \variable{Topology} information stored at the type level. Then, we also need to be able to create a graphical representation of a composed \variable{StateMachine} given a graphical representation of its sub-\variable{StateMachine}s. Depending on the used constructor, and on the graphical representation that we would like to obtain, we can decide how to represent the composed machine.

% =============================================================================
\section{Extending the architecture}

In real-life projects an application is rarely composed by only one aggregate, one policy, and one projection.
To understand how we can compose multiple aggregates, policies and projections together to model more complex workflows,
we can use the interface provided by \variable{Arrow} and \variable{ArrowChoice}, or, alternatively, the \variable{Profunctor}, \variable{Strong}, and \variable{Choice} type classes. Effectively, this means that the composition does not depend on our use of state machines; composition using another data type satisfying these type classes would also work in pretty much the same way.

Let's continue with our running example and consider that we would like to automatically start the delivery process once a payment is completed. As a first step, we
introduce a new aggregate that we implement as a ``Basic'' state machine:
\[
\begin{array}{@{\hspace{0mm}}l}
  \keydata~\mathit{ShippingCommand} = \ldots\\
  \\
  \keydata~\mathit{ShippingEvent} = \ldots\\
  \\
  \mathit{shipping} :: \mathit{StateMachine}~\mathit{ShippingCommand}~[\mathit{ShippingEvent}]\\
  \mathit{shipping} = \mathit{Basic} \ldots\\
\end{array}
\]
To connect this new aggregate with the rest of the application, we want to execute it as an alternative to the current write model, meaning that we want to handle either a \command{CartCommand} or a \command{ShippingCommand} and route it to the appropriate machine. We can achieve this using the ($\tripleplus$) operator from \variable{ArrowChoice} (or, equivalently, \variable{splitChoice} from the \variable{Choice} type class).
\[
\begin{array}{l}
  \mathit{writeModelWithShipping} :: \mathit{StateMachine} \\
    \quad (\mathit{Either}~\mathit{CartCommand}~\mathit{ShippingCommand})\\
    \quad [\mathit{Either}~\mathit{CartEvent}~\mathit{ShippingEvent}] \\
  \mathit{writeModelWithShipping} = \mathit{rmap} \\
    \quad (\mathit{fmap}~\mathit{Left} \triplebar \mathit{fmap}~\mathit{Right}) \\
    \quad (\mathit{writeModel} \tripleplus \mathit{shipping}) \\
\end{array}
\]
Now we have a way to process both commands concerning the cart payment and the shipping, but there is still no automatic connection between the two components. We would like to connect them by specifying that whenever a payment is completed the shipping process must start. The word \emph{``whenever''} indicates that we need to use a policy, specifically one that consumes \event{CartEvent}s and produces \command{ShippingCommand}s.
\[
\begin{array}{l@{\hspace{1mm}}c@{\hspace{1mm}}l}
  \multicolumn{3}{l}{\mathit{paymentCompletePolicy}} \\
  \multicolumn{3}{l}{\quad :: \mathit{StateMachine}~\mathit{CartEvent}~[\mathit{ShippingCommand}]} \\
  \multicolumn{3}{l}{\mathit{paymentCompletePolicy} = \mathit{stateless} ~\$~ \backslash\keycase} \\
    \quad\mathit{CartPaymentInitiated} & \to & [] \\
    \quad\mathit{CartPaymentCompleted} & \to & [\mathit{StartShipping}] \\
\end{array}
\]
We now need to connect our new \policy{paymentCompletePolicy} policy to our \variable{writeModelWithShipping} machine. Policies are always connected via the \variable{Feedback} constructor of the \variable{StateMachine} data type. To use it, we need to align the types by ignoring some inputs and enlarging the return type:
\[
\begin{array}{l}
  \mathit{writeModelWithShipping'} :: \mathit{StateMachine} \\
    \quad (\mathit{Either}~\mathit{CartCommand}~\mathit{ShippingCommand}) \\
    \quad [\mathit{Either}~\mathit{CartEvent}~\mathit{ShippingEvent}] \\
  \mathit{writeModelWithShipping'} = \\
    \quad \mathit{Feedback}~\mathit{writeModelWithShipping}~\$ \\
    \qquad \mathit{rmap}~(\mathit{fmap}~\mathit{Right})~\mathit{paymentCompletePolicy} \\
    \qquad\quad \triplebar~\mathit{stateless}~(\mathit{const}~[]) \\
\end{array}
\]
Next, we need another projection to provide the user with some queryable information about the status of the shipping. It is a process which consumes \event{ShippingEvent}s and produces some new \view{ShippingInfo} value:
\[
\mathit{shippingInfo} :: \mathit{StateMachine}~\mathit{ShippingEvent}~[\mathit{ShippingInfo}]
\]
The only thing that remains to be done is connecting the machine (\variable{writeModelWithShipping'}) and the two projections (\projection{paymentStatus} and \projection{shippingInfo}). We first pair up the projections to set up our read model and then we link it to our write model in a sequential fashion:
\[
\begin{array}{l}
  \mathit{readModel} :: \mathit{StateMachine} \\
    \quad (\mathit{Either}~\mathit{CartEvent}~\mathit{ShippingEvent}) \\
    \quad [\mathit{Either}~\mathit{CartView}~\mathit{ShippingInfo}] \\
  \mathit{readModel} = \mathit{rmap} \\
    \quad (\mathit{fmap}~\mathit{Left}~\triplebar~\mathit{fmap}~\mathit{Right}) \\
    \quad (\mathit{paymentStatus} \tripleplus \mathit{shippingInfo}) \\
  \\
  \mathit{cartAndShipping} :: \mathit{StateMachine} \\
    \quad (\mathit{Either}~\mathit{CartCommand}~\mathit{ShippingCommand}) \\
    \quad [\mathit{Either}~\mathit{CartView}~\mathit{ShippingInfo}] \\
  \mathit{cartAndShipping} = \\
    \quad \mathit{Kleisli}~\mathit{writeModelWithShipping'}~\mathit{readModel} \\
\end{array}
\]

% =============================================================================
\section{Future work}

Some possible directions for future work on the subject could be:

\begin{itemize}
  \item
    The \variable{Feedback} constructor of the \variable{StateMachine} data type is used to implement looping behaviour. Classically, such behaviour is implemented in terms of the \variable{ArrowLoop} or \variable{Costrong} type classes. It would be interesting to investigate the relation between \variable{Feedback} and such type classes, potentially removing the \variable{Feedback} constructor from the \variable{StateMachine} data type and replacing it with a constructor more similar to the \variable{loop} or \variable{unfirst} functions coming from the aforementioned type classes.
  \item
    The \variable{Sequential} and \variable{Kleisli} constructors both deal with sequential composition of state machines. It would be interesting to find a way to unify them.
  \item
    With the current implementation, we are tracking at the type level only a list of the allowed state transitions. There is no connection between the allowed transitions and the inputs which trigger them. In other words, we are not restricting in any way the inputs that we could receive in a given state. It would be interesting to implement a way to express that in a certain state only certain inputs are valid.

    This would allow stating explicitly which are the expected inputs in a given state when implementing a state machine, avoiding the need to implement error handling for such unwanted cases.
  \item
    The current implementation of \Crem does not take advantage of any concurrency or parallelism. In cases where a state machine is actually composed of multiple sub-machines in parallel, it could be interesting to allow running them on separate threads to improve the execution time.
  \item
    In the spirit of parallelising as much as possible a given machine, or anyway optimizing it in other ways, it could make sense to introduce an optimization step which would restructure the abstract syntax tree which constitutes a \variable{StateMachine}. Simple optimizations could be given by the algebraic structure provided by the categorial structure; for example any identity machine composed sequentially could be removed; or it would be possible to use the distributive law between sequential and parallel composition to improve the parallelisability of a machine.
\end{itemize}

Another potential area of future development concerns how to test systems built with \Crem and/or with the architecture described in the \emph{Domain-Driven Design} section.

Some potential aspects to investigate in that area could be:

\begin{itemize}
  \item
    Using a compositional library like \Crem, the Domain-Driven Design architecture described above could be implemented as a single state machine receiving commands as inputs and emitting views as outputs.

    One way to test such a system would be to proceed with unit testing, treating the whole domain as a unit, feeding commands to it and asserting the expected state on the view observed by the user.
  \item
    Another way to test such a system would be to use property-based testing, generating commands randomly and making assertions on the invariants of the views.

    A potential direction on investigation could be using something like linear temporal logic to provide a language to express system invariants, along the lines to what has been done with Quickstrom~\citep{Quickstrom}.
  \item
    Pushing it even further, we can notice that our domain and the user (in fact, any external system interacting with our system) together form a cycle. Therefore, if we implement our user as a state machine (potentially an effectful or probabilistic one) receiving views and producing commands, we could close the loop and let our application run as long as we like. At this point we could use this new complete system to test the invariants of the domain.
  \item
    Another way in which such a system could potentially be tested would be to use languages like TLA+ or Alloy. A model could be exported using the information stored in the topology of the machine. Then it could be tested and verified using the chosen specification language.

\end{itemize}

% =============================================================================
\section{Conclusion}

We believe that compositionality and representability constitutes two extremely important aspects for the success of a software architecture. The former because it allows tackling simpler problems and then composing back their solutions to obtain solutions for more complex problems. The latter because it allows an easier interaction between business and domain experts and software developers.

Crem makes it easy to keep the graphical representation synchronised with the implementation, providing a reliable source of graphical documentation.

Inspired by Domain-Driven Design and thanks to the Haskell type system we were able to create the \Crem library, which allows architecting systems in a composable and representable way, without sacrificing the developer experience while implementing such systems.

Moreover, the architecture we propose helps to clarify the role of the various components commonly used in a Domain-Driven Design architecture, describing more precisely which is their role inside the domain.

\balance

% =============================================================================
%%
%% The acknowledgments section is defined using the "acks" environment
%% (and NOT an unnumbered section). This ensures the proper
%% identification of the section in the article metadata, and the
%% consistent spelling of the heading.
\begin{acks}
We would like to thank the reviewers of FUNARCH '23 for their constructive comments, and especially Michael Sperber for shepherding the editing process; our paper is all the better for their thorough feedback and suggestions.

The initial development of the \Crem library happened while both authors were employed by Tweag.

The first author would also like to thank his former colleagues Richard Eisenberg, Alexis King, Sjoerd Visscher, Alexander Esgen, Nicolas Frisby and Daniele Palombi for the fruitful discussions and the feedback on the \Crem library. Moreover, he would like to thank his former colleague Alexei Drake for the fruitful collaboration in setting up a development environment for \Crem using \emph{Nix}.
\end{acks}

%%
%% The next two lines define the bibliography style to be used, and
%% the bibliography file.
\bibliographystyle{ACM-Reference-Format}
\bibliography{references}

\end{document}